\documentstyle[12pt,aaspp4]{article}   

\begin{document}

%
%
%
\newcommand{\app}{$\sim$}
\newcommand{\prop}{$\propto$}
\newcommand{\tento}[1]{$10^{#1}$}
\newcommand{\vzs}{$\times$}
\newcommand{\lb}{$\lambda$}
\newcommand{\mm}{$\pm$}
%
%
\newcommand{\hbc}{H$_\circ$}
\newcommand{\lsol}{$\cal L_\odot$}
\newcommand{\msol}{$\cal M_\odot$}
\newcommand{\ebv}{$E(\bv)$}
%
%
\newcommand{\kms}{km\,s$^{-1}$}
\newcommand{\ergs}{ergs\,s$^{-1}$}
\newcommand{\ergcs}{ergs\,cm$^{-2}$\,s$^{-1}$}
\newcommand{\ergcsa}{ergs\,cm$^{-2}$\,s$^{-1}$\,$\AA^{-1}$}
\newcommand{\cmc}{cm$^{-3}$}
%
%
\newcommand{\caiik}{\ion{Ca}{2}\,K\,\lb3933}
\newcommand{\cn}{CN\,\lb4200}
\newcommand{\gband}{CH\,Gband\,\lb4301}
\newcommand{\mgi}{\ion{Mg}{1}}
\newcommand{\mgb}{\ion{Mg}{1}+MgH\,\lb5175}
%
%
\newcommand{\brgama}{Br$\gamma$}
\newcommand{\pabeta}{Pa$\beta$}
\newcommand{\palfa}{Pa$\alpha$}
\newcommand{\halfa}{H$\alpha$}
\newcommand{\hbeta}{H$\beta$}
\newcommand{\hgama}{H$\gamma$}
\newcommand{\hdelta}{H$\delta$}
\newcommand{\heps}{H$\epsilon$}
\newcommand{\lya}{Ly$\alpha$}
%
%
\newcommand{\heii}{\ion{He}{2}\,\lb4686}
\newcommand{\hen}{\ion{He}{1}\,\lb5876}
\newcommand{\feii}{\ion{Fe}{2}}
%
%
\newcommand{\nif}{[\ion{N}{1}]\,\lb5199}
\newcommand{\nid}{[\ion{N}{2}]\,\lb6548,6584}
\newcommand{\oii}{[\ion{O}{2}]\,\lb3727}
\newcommand{\oiii}{[\ion{O}{3}]\,\lb5007}
\newcommand{\sii}{[\ion{S}{2}]\,\lb6717,6732}
\newcommand{\neiiia}{[\ion{Ne}{3}]\,\lb3869}
%
%
\newcommand{\oiiit}{[\ion{O}{3}]\,\lb}
\newcommand{\neiiit}{[\ion{Ne}{3}]\,\lb}
%
%
\newcommand{\heuv}{\ion{He}{2}\,\lb1640 + \ion{O}{3}]\,\lb1663}
\newcommand{\civ}{\ion{C}{4}\,\lb1549}
%

%
\newcommand{\nfofo}{NGC\,4151}
%

%
%

\title{The Narrow Line Region of NGC\,4151: a Turbulent
Cauldron\footnote{Based on observations made with the NASA/ESA {\it
Hubble Space Telescope}, obtained at the Space Telescope Science
Institute, which is operated by the Association of Universities for
Research in Astronomy, Inc., under NASA contract NAS\,5-26555.}}

\author{Cl\'audia Winge\,\altaffilmark{2}, David J. Axon\,\altaffilmark{3,4}, 
F.D. Macchetto\,\altaffilmark{3}}
\affil{Space Telescope Science Institute, 3700 San Martin Drive, Baltimore, 
MD21218, USA}

\and
\author{A. Capetti}
\affil{Scuola Internazionale Superiore di Studi Avanzati, Via Beirut 2-4, 
34014 Trieste, Italy}

\altaffiltext{2}{CNPq Fellowship, Brazil; winge@stsci.edu}

\altaffiltext{3}{Affiliated to the Astrophysics Division of ESA}

\altaffiltext{4}{Nuffield Radio Astronomy Laboratories, Jodrell Bank, 
Macclesfield, Cheshire SK11 9DL, UK}

\begin{abstract}

We present the first results of the Hubble Space Telescope/Faint Object
Camera  long-slit spectroscopy of the inner 8\arcsec\ of the Narrow
Line Region of \nfofo\ at a spatial resolution of 0\arcsec.029. The
emission gas is characterized by an underlying general orderly
behaviour, consistent with galactic rotation, over which are superposed
kinematically distinct and strongly localized emission structures. High
velocity components shifted up to \app\ 1500 \kms\ from the systemic
velocity are seen, associated with individual clouds located
preferentially along the edges of the radio knots. Off-nuclear blue
continuum emission is also observed, associated with the brightest
emission line clouds. Emission line ratios like \neiiia/\oii, and
\oii/\hbeta\ are observed to vary substantially between individual
clouds. We advance the general picture that, as in other Seyfert
galaxies observed with HST (e.g. NGC\,1068, Mrk\,573), the interaction
of the radio jet with the ambient gas strongly influences both the
morphology and the physical conditions of the NLR.

\end{abstract}


\section{Introduction}

Ground-based long-slit spectroscopy of the Extended Narrow Line Region
(ENLR -- a few kpc scale) of the Seyfert 1 galaxy \nfofo\ has shown
that the kinematics, ionization structure and morphology of the
emission-line gas are consistent with photoionization of gas in the
galactic disk by an anisotropic radiation field produced by the active
nucleus (Penston et al.  \markcite{praetal90} 1990). The situation on
the scale of the Narrow Line Region (sub-kpc) is however, much less
clear as there is a large misalignement between the ENLR at PA
\app\ 50\arcdeg\ (P\'erez et al.  \markcite{pgtetal89} 1989; Robinson
et al. \markcite{rvaetal94} 1994) and the NLR which is co-spatial with
the radio jet at PA \app\ 77\arcdeg\ (Pedlar et al.
\markcite{pkletal93} 1993). If both the NLR and ENLR structures are to
be explained in terms of an ``ionization cone'' then this misalignment
implies either a very broad (opening angle \app\ 120\arcdeg) cone of
ionizing radiation, intercepting the disk at grazing incidence (Pedlar
et al. \markcite{phau92} 1992, Vila-Vilar\'o \markcite{bv-v93} 1993,
Robinson et al.  \markcite{rvaetal94} 1994) or a geometry which is
distinct from the ``standard'' unified model which has the cone axis
fixed to that of the radio ejecta (Evans et al.  \markcite{etketal93}
1993).

The complex structure of clouds and filaments in the inner 5\arcsec\ of
the NLR revealed by {\it Hubble Space Telescope} (HST) images  suggests
that interaction with the radio jet may have an important influence on
the observed morphology.  In this Letter, we present the first results
of new HST long-slit spectroscopy of the central 8\arcsec\ of
\nfofo\ at a spatial resolution of 0\arcsec.029 which enables us to
separate for the first time the emission from the individual NLR
clouds, and demonstrates that the interaction with the radio ejecta
strongly affects the local kinematic and ionization conditions of the
emission gas. Our results imply that it is this interaction, rather
than  anisotropic illumination, which determines the basic structure of
the NLR.

For a distance to \nfofo\ of 13.3 Mpc, 0\arcsec.1 corresponds to a 
linear scale of 6.4 pc in the plane of the sky.

\section{Observations and Results}

The spectroscopic data consist of four spectra obtained with the Faint
Object Camera (FOC) f/48 long-slit facility on board the HST. The slit
width is 0\arcsec.063, and its useful length about 10\arcsec. We have
also used FOC f/96 images with filters centered at the
\oiii\ emission-line and nearby continuum. Table \ref{log_tab}
summarizes the relevant information on the above data set.  The f/96
images were reduced using the standard pipeline procedure, as described
in the FOC Handbook (Nota et al. \markcite{netal96} 1996).

The spectra were obtained with the slit at PA\,=\,47\arcdeg, and cover
the interval 3650 -- 5470 \AA. The data reduction followed the same
procedure as described in Macchetto et al. \markcite{mmaetal97} (1997);
wavelength resolution is 5.1 \AA\ \app\ 300 \kms\ at \oiii, and the
spatial PSF is of the order of 0\arcsec.08; flux calibration was
obtained using the IUE standard LDS749B. By comparing the brightness
profile of the \oiii\ emission line on the spectra with that on the
f/96 image, we were able to pinpoint the location of the slit to within
0\arcsec.06 (2 pixels at the f/48 resolution). Figure \ref{plate1_fig}
(Plate \ref{plate1_fig}) shows the slit positions superimposed on the
\oiii\ image. The spectra are located on the nucleus, and 0\arcsec.23,
0\arcsec.40 and 0\arcsec.46 NW of it, and will be hereafter referred to
as POS\,1, POS\,2, POS\,3 and POS\,4, respectively. The nuclear
spectrum was heavily saturated in the central 15 pixels, due to the
strong continuum emission. The presence of a complex radial velocity
structure is evident in all the brightest narrow emission lines, such
as [\ion{O}{2}], [\ion{Ne}{3}], and [\ion{O}{3}]. To study the detailed
kinematics of the gas the \oiiit4959,5007 emission lines, which provide
the best resolution and S/N ratio, were extracted in spatial windows
varying from two to 20 pixels along the slit, and both lines fitted
simultaneously using two to six Gaussian components, constraining the
physical parameters of the corresponding components of each line to
their theoretical values (3:1 intensity ratio, 48 \AA\ separation, and
same FWHM).

Figure \ref{plate2_fig} (Plate \ref{plate2_fig}) shows the combined
VLA$+$MERLIN 5 GHz radio map from Pedlar et al. \markcite{pkletal93}
(1993), overlaid on the \oiii\ image of \nfofo.  The registration was
made assuming that the optical nucleus corresponds to component C4, as
advocated by Mundell et al.\markcite{mpbetal95} (1995). There is an
apparent anti-correlation between the optical and the radio emission,
with the brightest emission-line filaments surrounding the radio knots
as would be expected if the radio-jet is clearing a channel in the
surrounding medium and enhancing the line emission along its edges by
compression of the ambient gas.

\section{Kinematics}

Figure \ref{rotcurv_fig} shows the results of the \oiiit4959,5007
fitting for three of the four slit positions (the POS\,3 and POS\,4
data are very similar to each other and do not give independent
information). The rotation curves were constructed assuming that the
narrowest, closest to the systemic velocity, and usually the brightest
component (filled squares) represent the ``main'' NLR gas emission. The
curves are plotted as a function to the distance to the center of the
slit, defined as a line passing through the nucleus, perpendicular to
the slit direction. The solid line on each panel represents the
integrated \oiii\ spatial profile.

The general behavior of the main component is consistent with normal
galactic rotation as defined by the ENLR gas emission (Robinson et al.
\markcite{rvaetal94} 1994), with strong but localized perturbations
superimposed, represented by one, or sometimes two, narrow (FWHM $<$
800 \kms) secondary components either blue- or redshifted relative to
the core of the line (shown as open circles and squares in Figure
\ref{rotcurv_fig}). In the inner 2\arcsec\ of the  nuclear spectrum,
the behaviour of the emission gas shows a strong deviation from the
outermost rotation curve.

It is important to note that the most kinematically disturbed features
are found to be associated with the emission clouds located along the
edges of the strongest radio emission components: the red/blue-shifted
knots at \app\ 1\arcsec\ SW of the center of the slit in the POS\,2
spectrum are located at one side of  the C2 radio component, and are
well fitted by two Gaussian components centered at $+$1300 and $-$800
\kms\ from the core velocity, and FWHM \app\ 330 and 490 \kms,
respectively; the high velocity ($\Delta$v \app\ $-$1700 \kms) cloud
{\bf (b)} in POS\,3 lies also alongside this same radio component; the
second-brightest radio component C3 is in front of the bright emission
complex located 0\arcsec.2--0\arcsec.5\ NW of the nucleus, where we
observe a clear double-peaked profile in all emission lines, well
represented by two narrow (FWHM \app\ 430\kms) components, one of which
is blueshifted by \app\ 660 \kms\ from the core velocity.

All these features are characterized by a very small projected size:
the two high-velocity clouds in POS\,2 discussed above are 0\arcsec.17
to 0\arcsec.23 (less than 15 pc) across, while the double-peaked
feature in POS\,3 (see Figure \ref{sampspec_fig}c) is resolved along a
0\arcsec.3 (20 pc) bin, giving further support to the idea that the NLR
morphology is strongly shaped in walls and voids by the expanding,
fast-moving radio components.

\section{Physical Conditions of the Emission Gas}

If the localized kinematics perturbations described above are a
consequence of fast shocks created by interactions with ejected radio
material the physical conditions in the clouds may also be influenced
by the shocks.  Not only can the shocks compression lead to local
density enhancements (Taylor, Dyson \& Axon \markcite{tda92} 1992) but
these may also be accompanied by secondary photoionizing radiation
(Sutherland, Bicknell \& Dopita \markcite{sbd93} 1993; Morse, Raymond
\& Wilson \markcite{mrw96} 1996).

We clearly see that the emission-line ratios vary substantially on
scales of a few tenths of an arcsec, indicating that the density and
ionization state of the emission gas are strongly influenced by the
local conditions. We also detect several examples of off-nuclear
blue-continuum emission, consistently associated with the brightest
emission-line knots.

Figure \ref{sampspec_fig} presents a sample of the spectra observed,
corresponding to the regions {\bf (a)} to {\bf (c)} in Figure
\ref{rotcurv_fig}. The observed line intensities, relative to \hbeta\ =
1.0, are given in Table \ref{flux_tab}. As a reference we included the
``average'' NLR line intensities, obtained by adding up the spatially
integrated emission of the four slit positions, and the ground-based
data for the ENLR ($r > 6$\arcsec) from Penston et al.
\markcite{praetal90} (1990). Errors are of the order of 0.05 or smaller
for the line ratios, and 0.002 or smaller for the
\oiiit4363/\oiiit5007$+$4959 ratio.

Spectrum \ref{sampspec_fig}(a) corresponds to the brightest \oiii\ knot
in POS\,2, one of the clouds associated with the radio component C3,
located at a projected distance of 0\arcsec.18 SW (12 pc) of the
nucleus. Spectrum \ref{sampspec_fig}(b) corresponds to the region with
the high velocity cloud located 1\arcsec\ SW (64 pc) of the center of
POS\,3. This component can be clearly seen in the [\ion{O}{3}] profile.
The FWHM  of the ``main'' \oiii\ emission in this spectrum is
unresolved (310 \kms), while spectra from neighboring regions (6
pixels, 0\arcsec.17 \app\ 11 pc either side) show a \lb5007 line twice
as broad. The \hgama/\hbeta\ ratios in these two positions are
consistent with little or no reddening. All line ratios for Spectrum
\ref{sampspec_fig}(b) are for the core component only, and do not
include the high velocity cloud contribution.

Both of these spectra present similar ratios \neiiia/\hbeta, while
[\ion{Ne}{3}]/\oii\ in {\bf (a)} is about twice that in {\bf (b)}. The
high value of the \lb3869/\lb3727 ratio is a local characteristic of
region {\bf (a)} that is not observed in nearby  clouds, and there is
no clear evidence of kinematic perturbations in this spectrum.  The
fact that \heii\ in {\bf (a)} is not significantly stronger than in
{\bf (b)} points against the hypothesis of a matter-bounded cloud,
which will enhance the [\ion{Ne}{3}] emission relative to
low-ionization lines (Binette, Wilson \& Storchi-Bergmann
\markcite{bws96} 1996). It is also possible that the interaction with
the radio-jet would compress the gas, and thus lower the ionizaton
parameter, favoring the emission of low-ionization lines such as
[\ion{O}{2}] in region {\bf (b)}. Notice, however, that the
\lb3869/\lb3727 ratio for this region is similar to the ratio averaged
over all the observed area, where one would expect the effects of local
density enhancements to have been washed out.

For these two spectra we also measured ratios $R_{[O\,{\sc iii}]}
\equiv $ \oiiit4363/\oiiit5007$+$4959 \app\ 0.012 and 0.016,
respectively. Using the 3-level atom of McCall \markcite{mm84} (1984),
for a reasonable electron density range of \tento{3} to \tento{5}
\cmc\ these values correspond to [\ion{O}{3}] temperatures $T_{[O\,{\sc
iii}]}$ in the range 12,000 to 16,000 K, slightly above the limit of
$T_e <$ 11,000 K predicted by most photoionization models, but still
compatible with the value found for the purportedly photoionized,
kinematically undisturbed ENLR. In fact, the overall similarity of the
NLR line ratios to those of the ENLR suggests that variations in
electron denstities created by different degrees in shock compression
in individual clouds are probably the main cause of the observed
variations of ionization from knot to knot.

However, this is not the whole story as shown in Spectrum
\ref{sampspec_fig}(c) which, in addition to double-peaked profiles,
shows clear evidence of a localized blue continuum. Such localized
regions of off-nuclear blue continuum emission are seen in several
positions in our spectra, and are always associated to, but not exactly
coincident with, an emission-line knot. As we mentioned above, this is
a characteristic spectral signature of the free-free emission of the
hot gas created by the fast shocks  and therefore provides concrete
evidence for the existence of a local ionizing continuum which might be
playing a significant role in the ionization structure of the
individual clouds.

\section{Conclusions}

The high spatial resolution long-slit spectroscopy of the nuclear
region of \nfofo\ presented in this paper clearly show that the popular
picture of anisotropic illumination by the nuclear source is overly
simplistic to explain the complex morphology and  kinematics observed
in the NLR. Comparison of the radio and optical data shows that the
line emission is enhanced along the edges of the radio knots, as would
be expected to result from the interaction between the jet and the
surrounding medium, similar to what is observed for NGC\,1068
(Gallimore et al. \markcite{gbop96} 1996; Capetti, Macchetto \&
Lattanzi \markcite{cml97} 1997; Capetti, Axon \& Macchetto
\markcite{cam97} 1997).

The general behaviour of the \oiii\ velocity field is consistent with
disk rotation, with local deviations caused by turbulence (probably
associated with the sweeping/expanding motion of the radio cocoon). We
detect several distinct and very localized high velocity ($-$1700 to
$+$1300 \kms), narrow (FWHM $<$ 600 \kms) emission components, both
blue and redshifted, confined in projected dimensions of up to
0\arcsec.3. We find no evidence for systematic red/blue-shifted
deviations as expected by the presence  of a generalized outflow, as
advocated by Schulz \markcite{hs90} (1990). Detailed modeling of the
inner $|r| \lesssim 4$\arcsec\ kinematics will be presented in a
forthcoming paper.

The data shows that the emission-line ratios can vary substantially
within a few tenths of arcsec, indicating that the physical conditions
of the emission gas are also strongly influenced by the local
conditions. Furthermore, we detect off-nuclear blue continuum emission,
consistently associated with the brightest emission-line knots.

The clear morphological association between the radio and the optical
emission,  and the highly disturbed kinematical components in the NLR
of \nfofo\ revealed by our data imply that interaction with the radio
plasma rather than illumination shapes the NLR.  Within the context of
the present data, the relative balance between the two possible
explanations of the ionization structure variations we have discussed,
density fluctuations and/or local ionization, cannot be resolved as we
do not have density measurements of the individual clouds.
Nevertheless, it is evident that shocks associated with the jet-cloud
interactions play an important role in determining the physical
conditions of the {\it individual} emission clouds (Taylor, Dyson \&
Axon \markcite{tda92} 1992; Sutherland at al.\markcite{sbd93} 1993).

In order to unambiguously determine if local ionization effects play a
significant role in modifying the ionization structure of individual
clouds high quality measurements of the electron density in the clouds
need to be obtained from future HST spectroscopy.

\acknowledgments We thank the referee for carefully reading the
manuscript and making a number of constructive comments which improved
the clarity of the paper.

\newpage


\newpage


\figcaption[n4151plt1.ps]{[\ion{O}{3}] FOC f/96 image showing the
position of the spectrograph slit for the spectra listed in Table
\ref{log_tab}. The white segments represent the position of the spectra
shown in Figure \ref{sampspec_fig}. The actual length of the slit is
longer than represented. North is at the top and East is to the left.
\label{plate1_fig} }

\figcaption[n4151plt2.ps]{Contours of the VLA$+$MERLIN 5\,GHz radio-map
and knot identification from Pedlar et al. \protect\markcite{pkletal93}
(1993) overlayed on the same image as in Figure
\protect\ref{plate1_fig}. The emission gas tends to trace the
boundaries of the radio-jet and knots. \label{plate2_fig} }

\figcaption[n4151fig3.ps]{[\ion{O}{3}] rotation curves for three of the
observed positions, superimposed on the integrated \oiii\ spatial
profile (solid line), and plotted as a function of the distance to the
center of the slit. The main kinematic component, corresponding to gas
rotating in the plane of disk, is represented by filled squares. The
open circles and squares correspond to secondary narrow and extra
components, respectively (see text for details). Velocity errors are
estimated from the Gaussian fitting procedure. Notice the steep
velocity gradient in the inner 2\arcsec\ of the nuclear spectrum.
Regions {\bf (a)} to {\bf (c)} correspond to the spectra shown in
Figure \protect\ref{sampspec_fig}. The scale to the right refers to the
integrated flux in the \oiii\ profile, and the saturated region in the
nuclear spectrum is marked as {\bf sat.}. \label{rotcurv_fig} }

\figcaption[n4151fig4.ps]{Spectra corresponding to regions {\bf (a)} to
{\bf (c)} in Figure \protect\ref{rotcurv_fig}. Top: region {\bf (a)},
the brightest line-emission knot in POS\,2, showing strong \oiiit4363
and \neiiia; Middle: region {\bf (b)}, containing the high-velocity
knot at 1\arcsec\ SW in the POS\,3 spectrum. The high velocity
blueshifted emission from the knot is clearly visible in both \lb5007
and \lb4959 lines; Bottom: region {\bf (c)}, revealing double-peaked
profiles in all the observed lines. The dotted line shows the same
spectrum in a five times expanded scale to enhance the underlying
continuum. \label{sampspec_fig}}

\newpage



\begin{deluxetable}{lcccr}
\tablecolumns{5}
\tablewidth{0 pt}
\tablecaption{Journal of Observations \label{log_tab}}
\tablehead{ \colhead{\nfofo} & \colhead{} & \colhead{HST} & \colhead{} & \colhead{Exp. time} \\     
\colhead{Observation} & \colhead{Date} & \colhead{Rootname} & \colhead{Filter} & \colhead{(s)} }
\startdata
Spectra & 1996 Jul 03 & X38I0102T & F305LP & 1247 \nl
        &             & X38I0108T & F305LP & 697 \nl
        &             & X38I0109T & F305LP & 697 \nl
        &             & X38I010AT & F305LP & 697 \nl
Images  & 1994 Dec 03 & X2740101T & F501N  & 2160 \nl
        &             & X2740102T & F4ND$+$F501N & 900 \nl
        &             & X2740108T & F550M  & 1340 \nl
\enddata
\end{deluxetable}
\clearpage


\begin{deluxetable}{lccccccc}
\tablecolumns{8}
\tablewidth{0 pt}
\tablecaption{Observed Line Ratios: \hbeta\ = 1.0 \label{flux_tab}}
\tablehead{ \colhead{} & \colhead{} & \colhead{} & \multicolumn{3}{c}{Spectrum (c)} & \colhead{} & \colhead{} \\
\cline{4-6} \\
\colhead{Line} & \colhead{Spectrum (a)} & \colhead{Spectrum (b)} & \colhead{Main} & \colhead{Blue} & \colhead{Sum} & \colhead{NLR\tablenotemark{a}} & \colhead{ENLR\tablenotemark{b}} }
\startdata
\oii                 & 1.54\phn & 3.33\phn & 2.09\phn & 1.21\phn & 1.79\phn & 2.56\phn & 2.44\phn \nl
\neiiia              & 1.24\phn & 1.25\phn & 0.91\phn & 0.82\phn & 0.88\phn & 1.07\phn & 1.0\phn\phn  \nl
\hgama               & 0.44\phn & 0.53\phn & 0.39\phn& 0.34\phn & 0.37\phn & 0.39\phn & 0.42\phn \nl
\oiiit4363           & 0.19\phn & 0.34\phn & 0.37\phn & 0.27\phn & 0.27\phn & 0.25\phn & 0.24\phn \nl
\heii                & 0.14\phn & 0.19\phn & 0.26\phn & 0.31\phn & 0.28\phn & 0.27\phn & 0.38\phn \nl
\oiii                & 11.9\phn\phn\phn & 10.8\tablenotemark{c}\phn\phn\phn & 15.2\phn\phn\phn & 10.8\phn\phn\phn & 13.7\phn\phn\phn & 12.5\phn\phn\phn & 12.9\phn\phn\phn \nl
\sidehead{Ratios}
\lb3727/\lb5007      & 0.13\phn  & 0.31\phn  & 0.14\phn  & 0.11\phn  & 0.13\phn & 0.21\phn & 0.19\phn \nl
\lb3869/\lb3727      & 0.81\phn  & 0.38\phn  & 0.44\phn  & 0.68\phn  & 0.49\phn & 0.42\phn & 0.41\phn \nl 
$R_{[O\,{\sc iii}]}$\tablenotemark{d} & 0.012 & 0.016 & 0.014 & 0.018 & 0.015 & 0.015 & 0.014 \nl
\enddata
\tablenotetext{a}{Integrated emission from the four slit positions shown in Figure \ref{plate1_fig}.}
\tablenotetext{b}{From Penston et al. \markcite{praetal90}(1990), corresponding to the region 6--20\arcsec\ SW of nucleus.}
\tablenotetext{c}{Value does not include the high velocity cloud.}
\tablenotetext{d}{$R_{[O\,{\sc iii}]} \equiv $  \oiiit4363/\oiiit5007$+$4959}
\end{deluxetable}
\clearpage

%
%

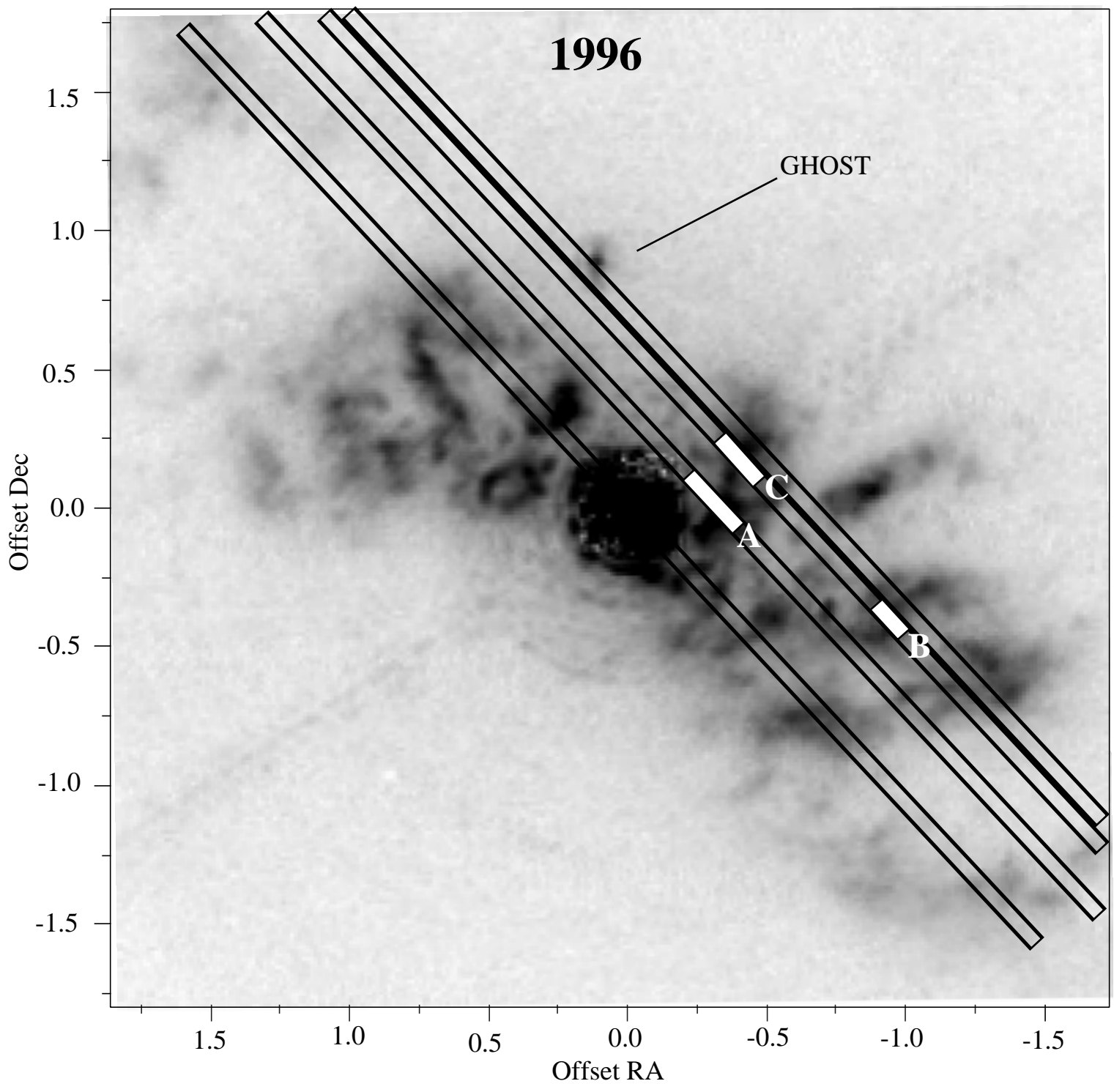
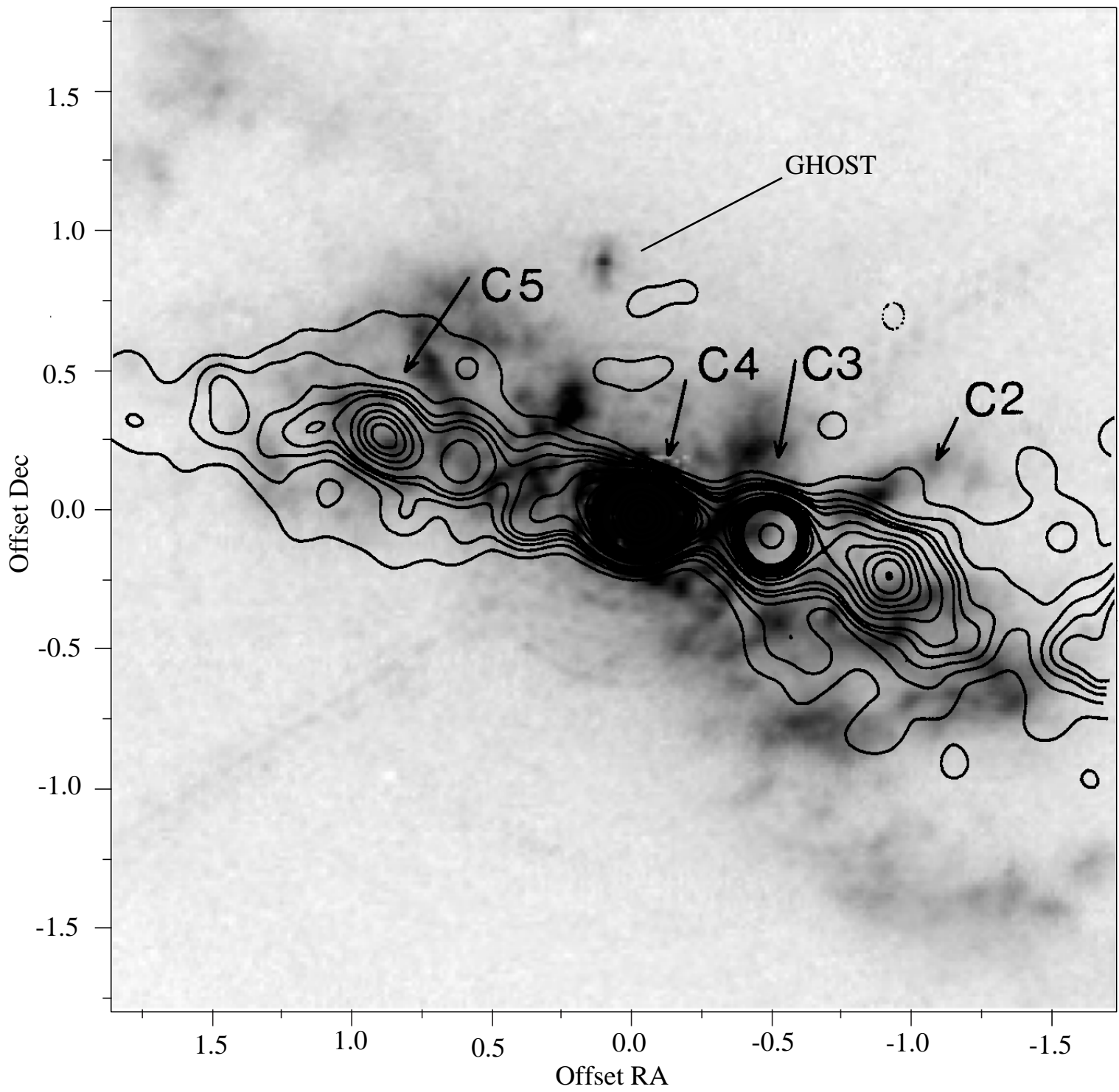
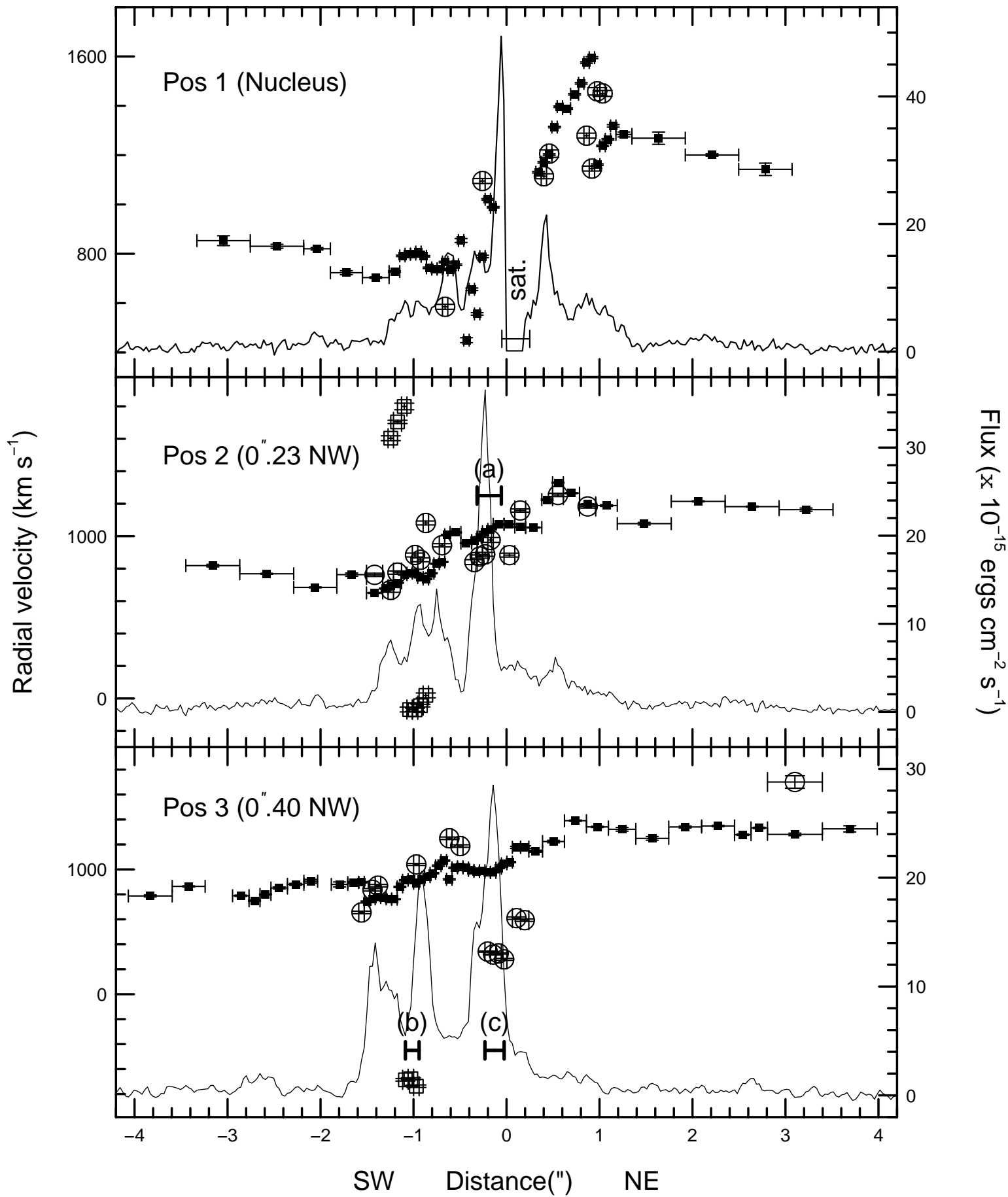
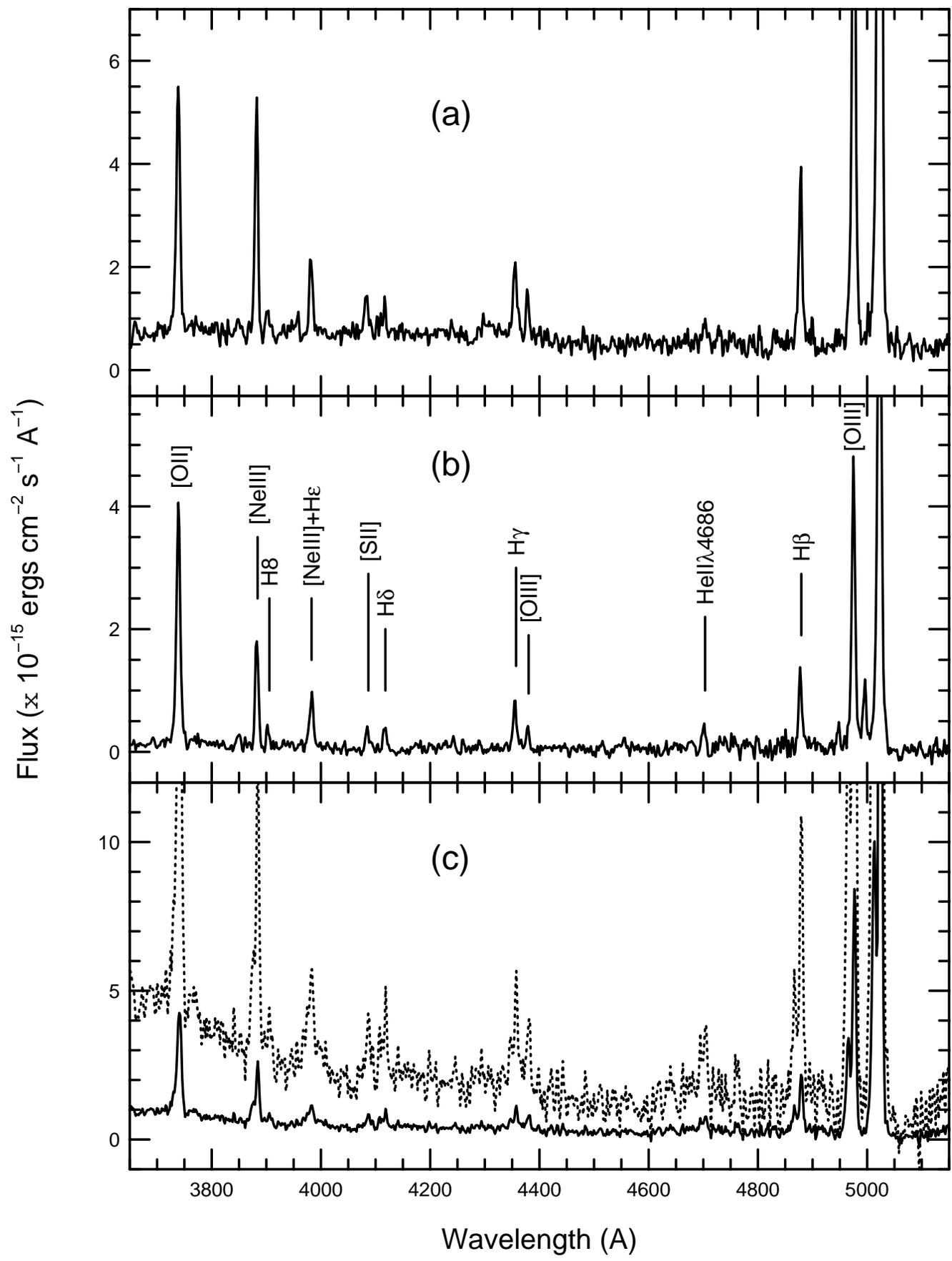
\end{document}